\documentclass[aps,pre,reprint,superscriptaddress,longbibliography]{revtex4-1}

\usepackage{graphicx}
\usepackage{times}
\usepackage{epstopdf}
\usepackage{amssymb}	% for math
\usepackage{amsfonts}
\usepackage{amsmath}
\usepackage{mathrsfs}
\usepackage{mathtools}
\usepackage{times}
\usepackage{xcolor}
\usepackage{adjustbox}

\newcommand{\thetas}{\boldsymbol{\theta}}
\newcommand{\zetas}{\boldsymbol{\zeta}}
\newcommand{\etas}{\boldsymbol{\eta}}
\newcommand{\ps}{\mathbf{p}}

\newcommand{\R}{\mathbb{R}}

\begin{document}

%\preprint{}

\title{Tensorial and bipartite block models\\
for link prediction in layered networks and temporal networks
}


%\author{}
%\thanks{}
%\affiliation{}

\author{Marc Tarr\'es-Deulofeu}
\email{marc.tarres@urv.cat}
\thanks{These authors contributed equally to the work}
\affiliation{Departament d'Enginyeria Qu\'{\i}mica, Universitat Rovira i Virgili, 43006 Tarragona, Catalonia}
\author{Antonia Godoy-Lorite}
\email{antonia.godoy@urv.cat}
\thanks{These authors contributed equally to the work}
%\affiliation{Departament d'Enginyeria Qu\'{\i}mica, Universitat Rovira i Virgili, 43006 Tarragona, Catalonia}
\affiliation{Department of Mathematics, Imperial College London, London SW7 2AZ, United Kingdom}
\author{Roger Guimer\`a}
\email{roger.guimera@urv.cat}
\thanks{Corresponding author}
\affiliation{Departament d'Enginyeria Qu\'{\i}mica, Universitat Rovira i Virgili, 43006 Tarragona, Catalonia}
\affiliation{ICREA, 08010 Barcelona, Catalonia}
\author{Marta Sales-Pardo}
\email{marta.sales@urv.cat}
%\thanks{Corresponding author}
\affiliation{Departament d'Enginyeria Qu\'{\i}mica, Universitat Rovira i Virgili, 43006 Tarragona, Catalonia}


\date{\today}

\begin{abstract}
	Many real-world complex systems are well represented as multilayer networks; predicting interactions in those systems is one of the most pressing problems in predictive network science. To address this challenge, we introduce two stochastic block models for multilayer and temporal networks; one of them uses nodes as its fundamental unit, whereas the other focuses on links. We also develop scalable algorithms for inferring the parameters of these models. Because our models describe all layers simultaneously, our approach takes full advantage of the information contained in the whole network when making predictions about any particular layer. We illustrate the potential of our approach by analyzing two empirical datasets---a temporal network of email communications, and a network of drug interactions for treating different cancer types. We find that modeling all layers simultaneously does result, in general, in more accurate link prediction. However, the most predictive model depends on the dataset under consideration; whereas the node-based model is more appropriate for predicting drug interactions, the link-based model is more appropriate for predicting email communication.

\end{abstract}

%\keywords{}

\maketitle

% -----------------------------------------------------------------------------------
% INTRODUCTION
% -----------------------------------------------------------------------------------

\section{Introduction}

Imagine a team of researchers looking for promising drug combinations to treat a specific cancer type for which current treatments are ineffective. The team has data on the effect of certain pairs of drugs on other cancer types, but the data are very sparse---only a few drug pairs have been tested on each cancer type, and each drug pair is tested in a few cancer types, at best, or has never been tested at all. The challenge is to select the most promising drug pairs for testing with the target cancer type, so as to minimize the cost associated to unsuccessful tests.

We can formalize this challenge as the following inference problem: We have a partial observation of the pairwise interactions between a set of nodes (drugs) in different ``network layers'' (cancer types), and we need to infer which are the unobserved interactions within each layer (drug interactions in each cancer type). This challenge is relevant for the many systems that can be represented as multilayer networks \cite{dedomenico13,bassett17,kefi16,pilosof17}, and is also formally analogous to the challenge of predicting the existence of interactions between nodes in time-resolved networks \cite{iribarren09,mucha10,gauvin14,peixoto15,ghasemian16,sapienza17,li17}. For instance, we would face the same situation if we had data about the daily e-mail or phone communications between users, and wanted to infer the existence of interactions between pairs of users on a certain unobserved day; in this case each layer would be a different day.

Here, we introduce new generative models that are suitable to address the challenge above. We model all layers concurrently, so that our approach takes full advantage of the information contained in all layers to make predictions for any one of them. Our approach relies on the fact that having information on the interactions in different layers aids the inference process; in other words, that the interactions in layers different from the one we are interested in are informative about the interactions in the query layer. For instance, biologically similar cancer types are likely to show similar responses to the same drug pairs, and similar days of the week (for instance weekdays versus weekends) are also likely to display similar communication patterns for pairs of users.
%
%Unlike previous approaches \cite{tiago,cris}, our models do not make any a priori assumptions about which layers are more informative about which other layers. Rather, this information is a byproduct of the inference process.

Our approach is based on recent results on probabilistic inference on stochastic block models, which has been successful at modeling the structure of complex networks \cite{guimera09,peixoto14,valles-catala17b} and at predicting the behavior in biological \cite{guimera13} and social \cite{roviraasenjo13,godoy-lorite16b} systems. In particular, we focus on mixed-membership stochastic block models \cite{airoldi08}, in which nodes are allowed to belong to multiple groups simultaneously. With these models it possible to model large complex networks with millions of links and, because they are more expressive than their fixed-membership counterparts, their predictive power is often superior \cite{godoy-lorite16b}.
%
%
%Unlike existing approaches, which do not take full advantage of the similarity between layers to aid in the inference process \cite{stanley16}, we introduce multi-layer models that naturally capture arbitrary similarities between layers.
%
We propose two different mixed-membership multi-layer network models---a tensorial model that takes nodes as the basic unit to describe interactions in different layers, and a bipartite model that takes links (or pairs of nodes) as the basic unit.
In our models, layers, as well as nodes or links, are grouped based on the similarities among the interaction patterns observed in them. This is in contrast to existing approaches, which do not take full advantage of the information that each layer carries about the structure of some other layers.

We illustrate our models and inference approaches by analyzing two datasets---a network of drug interactions in different cancer types, and a temporal network of email communications \cite{godoy-lorite16a}. We find that modeling all layers simultaneously, and assuming that they can be grouped, results in link predictions that are more accurate (in terms of standard metrics such as precision and recall) than those of single-layer models and of simpler multilayer models. However, the most predictive model (node-based or link-based) depends on the dataset under consideration. Indeed, whereas for drug interactions drug groups are very informative and, therefore, node-based models are most predictive, temporal email networks are best described in terms of links, that is, in terms of the relationships between pairs of individuals rather than the individuals themselves.

%The behavior of complex systems is strongly dependent on the underlying network of interactions between the systems components. The scientific community has put a lot of effort into the study of the structure of such network of interactions with the goal of not only understanding but also being able to predict the behaviour of these systems under different conditions. 
% -----------------------------------------------------------------------------------
% Models
% -----------------------------------------------------------------------------------

\section{Tensorial and bipartite mixed-membership block models for layered networks}

We aim to model $N$ nodes interacting by pairs in $M$ different layers; these layers correspond to the different contexts in which the nodes interact (for example, different cancer types or time windows). We represent these interactions as a layered graph $G$ whose links $(i,j,\ell)$ represent interactions between nodes $i$ and $j$ in layer (or at time) $\ell$. Moreover, we allow for multi-valued interactions so that $(i,j,\ell)$ can be of different types $r_{ij\ell}\in R$, where $R$ is a finite set. Note that we can use this formalism to model labels, attributes or ratings associated to the interactions \cite{guimera13,godoy-lorite16b}; graphs with binary interactions are therefore a particular case within this general framework in which $r_{ij\ell}=1$ if the interaction occurs and $r_{ij\ell}=0$ if it does not.

We consider two types of generative models---one that takes individual nodes as its basic unit, and one that models links (or node pairs). The first generative model, based on individual nodes, is as follows. There are $K$ groups of nodes and $L$ groups of layers. We assume that the probability that a node in group $\alpha$ has an interaction of type $r$ with a node in group $\beta$ in a layer in group $\gamma$ is $p_{\alpha \beta \gamma}(r)$. 
%
%For each trio of node group $\alpha$, node group $\beta$, and layer group $\gamma$ there is a probability distribution $p_{\alpha \beta \gamma}(r)$ over the interaction types $r\in R$; $p_{\alpha \beta \gamma}(r)$ is the probability ---> I have removed this because the probaility distribution part is only true if edge classes are exclusive
%
Furthermore, we assume that both nodes and layers can belong to more than one group. To model such mixed group memberships \cite{airoldi08}, to each node $i$ we assign a vector $\theta_i \in \R^K$, where $\theta_{i\alpha}\in [0,1]$ denotes the probability that node $i$ belongs to group $\alpha$. Similarly, to each layer $\ell$ we assign a vector $\eta_{\ell \gamma} \in \R^L$.
These vectors are normalized so that $\sum_\alpha \theta_{i\alpha} = \sum_\gamma \eta_{\ell \gamma} = 1$.
The probability that link $(i,j,\ell)$ is of type $r$ is then 
\begin{equation}
 \Pr[r_{ij\ell} = r] = \sum_{\alpha\beta\gamma} \theta_{i\alpha} \theta_{j\beta} \eta_{\ell\gamma} p_{\alpha\beta\gamma}(r) \, .
 \label{eq:node-model}
\end{equation}
Note that if link types are exclusive (i.e. each edge can be of only one type), the probability tensor must satisfy the constraint $\sum_{r \in R} p_{\alpha\beta\gamma}(r)=1$. 
Since this model is an extension of the mixed-membership stochastic block model \cite{airoldi08,godoy-lorite16b} where the probability matrices become tensors because of the multiple layers \cite{dedomenico13}, we call it the tensorial mixed-membership stochastic block model (T-MBM).

Our second generative model for layered networks is as follows. Instead of assuming that nodes belong to groups, we  assume that it is links (or pairs of nodes, rather than individual nodes) that belong to groups \cite{peixoto15}. In this model we have $J$ groups of links, and the probability that a link $e_{ij}\equiv e$ in group $\alpha$ is of type $r$ in a layer $\ell$ in group $\gamma$ is $p_{\alpha\gamma}(r)$. We also assume that links can belong to more than one group so that $\zeta_{e\alpha}$ is the probability that link $e$ belongs to group $\alpha$ and $\sum_{\alpha}\zeta_{e\alpha}=1$. As before, to each layer $\ell$ we also assign a vector $\eta_{\ell} \in \R^L$ of group memberships. Then, the probability that a given link in a particular layer is of type $r$ is
\begin{equation}
 \Pr[r_{ij\ell} = r] = \Pr[r_{e\ell} = r] = \sum_{\alpha\gamma} \zeta_{e\alpha} \eta_{\ell\gamma} p_{\alpha\gamma}(r) \, ,
 \label{eq:link-model}
\end{equation}
where, as before, if link types are exclusive the probability matrices satisfy the condition $\sum_{r \in R} p_{\alpha\gamma}(r)=1$.
This model can be seen as a bipartite model with two types of elements, links and layers. In this representation, a link $e_{ij}$ has a connection of type $r$ to a layer $\ell$ if $r_{ij\ell}=r_{e\ell}=r$. Therefore, we call this model the bipartite mixed-membership stochastic block model (B-MBM). 

These models are novel in a number of ways. First, unlike other models of multi-layer networks \cite{peixoto15,debacco17}, they do not assume any particular order in the layers, and therefore do not impose any restrictions to how layers should be grouped. This is in contrast to approaches for temporal networks that can only group layers corresponding to consecutive times. While such restriction simplifies the task of grouping layers, it also eliminates the possibility of identifying, for example, periodicities in temporal networks. More importantly, this restriction prevents models from being applicable to non-temporal multilayer networks. Our models eliminate this restriction.

Second, unlike other models \cite{stanley16,debacco17}, ours assume that group memberships (of nodes or links) do not change from layer to layer. Rather, we argue that in many relevant situations membership is determined by intrinsic properties of nodes or links and so long as these properties do not change, group membership should not change either. For example, membership of individuals to groups in social networks is related to demographic and socio-economic characteristics, which are unlikely to change in periods of months or even a few years.
%Likewise, the nature and frequency of social relationships do change over time, but not as rapidly as not to be considered stable for the duration of our datasets.
Or in drug-interaction networks, membership of drugs to groups is related to the mechanism of action and the targets of the drug \cite{guimera13}, which do not change regardless of the situation in which the drug is used.

Third, our models naturally deal with situations in which links have metadata, that is, situations in which nodes are not only connected or disconnected, but rather can be connected with links of different types.

Finally, unlike other models of multilayer and temporal networks \cite{peixoto15,debacco17,stanley16}, in our models nodes/links and layers do not belong to a single group, but rather to a mixture of groups \cite{airoldi08,godoy-lorite16b}. This allows us to develop efficient expectation-maximization algorithms that can be massively parallelized \cite{jefries17} and, at the same time, provide better predictions than single-group models \cite{godoy-lorite16b}. 

% -----------------------------------------------------------------------------------
% INFERENCE
% -----------------------------------------------------------------------------------

\section{Inference equations and expectation-maximization algorithms}

Given a set $G^O$ of observed links types, our goal is to predict the types $r_{ij\ell}$ of links $(i,j,\ell) \not\in G^O$ whose type is unknown. Because marginalizing over the parameters in our models (Eqs.~(\ref{eq:node-model}) and (\ref{eq:link-model})) is too time-consuming, here we present a maximum likelihood approach (and the corresponding expectation-maximization algorithms) for the two models above.

\subsection{Tensorial model}

Given the generative T-MBM model in Eq.~(\ref{eq:node-model}), and abbreviating its parameters as $\thetas, \etas, \ps$, the likelihood of the model is
\begin{equation}
P(G^O | \thetas, \etas, \ps) = \prod_{(ij\ell) \in G^O} \sum_{\alpha\beta\gamma} \theta_{i\alpha} \theta_{j\beta} \eta_{\ell\gamma} p_{\alpha\beta\gamma}(r_{ij\ell}) \, . 
\label{eq:likeli-t}
\end{equation}

As we show below (Appendix~\ref{app:tensorial}), the values of the parameters that maximize this likelihood satisfy the following equations
\begin{eqnarray}
\label{eq:update-theta-t}
\theta_{i\alpha} & = & \frac{\sum_{(j\ell) \in \partial i} \sum_{\beta\gamma} \omega_{ij\ell}(\alpha,\beta,\gamma)}{d_i} \, ,\\
\label{eq:update-eta-t}
\eta_{\ell\gamma} & = & \frac{\sum_{(ij) \in \partial \ell} \sum_{\alpha\beta} \omega_{ij\ell}(\alpha,\beta,\gamma)}{d_{\ell}} \, , \\
\label{eq:update-pr-t}
p_{\alpha\beta\gamma}(r) & = & \frac
{\sum_{(i,j,\ell) \in G^O | r_{ij\ell}=r} \omega_{ij\ell}(\alpha,\beta,\gamma)}
{\sum_{(i,j,\ell) \in G^O} \omega_{ij\ell}(\alpha,\beta,\gamma)} \, .
\end{eqnarray}
Here, $\partial i = \{ (j,\ell) | (i,j,\ell) \in G^O \}$ are the set of observed layer-specific neighbors of node $i$ and $d_i=|\partial i|$ is the total degree of the node in all the layers. Similarly, $\partial \ell = \{ (i,j) | (i,j,\ell) \in G^O \}$ is the set of observed links in layer $\ell$ and $d_\ell=|\partial \ell|$. Finally, $\omega_{ij\ell}(\alpha,\beta,\gamma)$ is the estimated probability that the type of a given link $r_{ij\ell}$ is due to $i$, $j$ and $\ell$ belonging to groups $\alpha$, $\beta$ and $\gamma$ respectively, and is given by 
\begin{equation}
\label{eq:update-omega}
\omega_{ij\ell}(\alpha,\beta,\gamma) 
= \frac{\theta_{i\alpha} \theta_{j\beta} \eta_{\ell \gamma} p_{\alpha \beta \gamma}(r_{ij\ell})}{\sum_{\alpha'\beta'\gamma'} \theta_{i\alpha'} \theta_{j\beta'} \eta_{\ell \gamma'} p_{\alpha' \beta' \gamma'}(r_{ij\ell})} \, .
\end{equation}

These equations can be solved iteratively with an expectation-maximization algorithm, starting with an initial estimate of $\thetas$, $\etas$, and $\ps$ and, then, repeating the following steps: (i) use Eq.~\eqref{eq:update-omega} to compute $\omega_{ij\ell}(\alpha, \beta, \gamma)$ for $(i,j,\ell) \in G^O$ (expectation step); (ii) use Eqs.~\eqref{eq:update-theta-t}-\eqref{eq:update-pr-t} to compute $\thetas$, $\etas$, and $\ps$ (maximization step).

\subsection{Bipartite model}

Similarly, the likelihood of the B-MBM is
\begin{equation}
P(G^O | \zetas, \etas, \ps) = \prod_{(e,\ell) \in G^O} \sum_{\alpha\gamma} \zeta_{e\alpha} \eta_{\ell\gamma} p_{\alpha\gamma}(r_{e\ell}) \, ,
\label{eq:links-likeli-b}
\end{equation}
and the maximum likelihood estimators of the parameters satisfy
\begin{eqnarray}
\label{eq:links-update-theta-b}
\zeta_{e\alpha} & = & \frac{\sum_{\ell \in \partial e} \sum_{\beta\gamma} \phi_{e\ell}(\alpha, \gamma)}{d_{e}} \, , \\
\label{eq:links-update-eta-b}
\eta_{\ell\gamma} & = & \frac{\sum_{e \in \partial \ell} \sum_{\alpha} \phi_{e\ell}(\alpha, \gamma)}{d_{\ell}} \, , \\
\label{eq:links-update-pr-b}
p_{\alpha\gamma}(r) & = & \frac
{\sum_{(e\ell) \in G^O | r_{e\ell}=r} \phi_{e\ell}(\alpha, \gamma)}
{\sum_{(e\ell) \in G^O} \phi_{e\ell}(\alpha, \gamma)} \, .
\end{eqnarray}
Here $\partial e = \{\ell | (e,\ell) \in G^O \}$ are the observations of link $e_{ij}$ in all layers and $d_e=|\partial e|$. As before, $\partial \ell = \{ e | (e,\ell) \in G^O \}$ are the observed links in layer $\ell$ and $d_\ell=|\partial \ell|$. Finally, $\phi_{e\ell}(\alpha, \gamma)$ is the estimated probability that the type of a specific link $r_{e\ell}$ is due to $e$ and $\ell$ belonging to groups $\alpha$ and $\gamma$ respectively; we can compute $\phi_{e\ell}(\alpha, \gamma)$ as
\begin{equation}
\label{eq:links-update-omega-b}
\phi_{e\ell}(\alpha, \gamma) 
= \frac{\zeta_{e\alpha} \eta_{\ell \gamma} p_{\alpha \gamma}(r_{e\ell})}{\sum_{\alpha'\gamma'} \zeta_{e\alpha'} \eta_{\ell \gamma'} p_{\alpha' \gamma'}(r_{e\ell})} \, .
\end{equation}

Like in the tensorial model, these equations can be solved iteratively using an expectation-maximization algorithm.

\section{Model comparison on real data}

\subsection{Datasets}

We perform experiments on two different datasets: the time-resolved email network of an organization spanning one year \cite{godoy-lorite16a}, and a network of drug-drug interactions in different cancer cell lines \cite{DREAM15}. In the email dataset, we represent each day as a different layer of the multi-layer network, and two users are considered to interact in a given day if they send at least one email in either direction during that day. We consider several e-mail networks that correspond to e-mail communications within organizational units (see Table \ref{tab:data}).

In the drug-drug interactions dataset, each layer corresponds to a different cancer cell line and we have information on the effects of some drug pair combinations on some cancer cell lines \cite{DREAM15}. In contrast to the email dataset, in which all the interactions (or lack of interaction) are observed, this dataset is sparsely observed---we have information about 1.5\% of the drug pairs. Specifically, the available experimental data is a real-valued magnitude representing the combined efficiency of two drugs on a particular cell line. These magnitudes range from large absolute values, in which case the interaction is said to be synergistic (if it is positive) or antagonistic (if it is negative), to small absolute values, in which case the interaction it is said to be additive. In an additive interaction, the application of the two drugs together has an efficiency equal or similar to the sum of the efficiencies of each drug administered separately. By contrast, in a synergistic (antagonistic) interaction the efficiency of the two drugs administered together is significantly higher (lower) than the sum of the efficiencies of each drug administered separately.
%In order to divide the continuous values into three categories (synergistic, additive and antagonistic), we set two thresholds at -20.0 and 20.0, so that interactions with a score lower than -20.0 are classified as antagonistic, those with a score higher than 20.0 are classified as synergistic, and those in between are considered additive \cite{DREAM15}.

In Table~\ref{tab:data} we show the characteristics of each dataset in terms of the types of links $R$, the total number of nodes, the total number of layers, the total number of possible links, and the number and fraction of actually observed links. In all cases, we fitted and validated our models using a 5-fold cross-validation scheme (Appendix~\ref{app:details}).

%We divided the data into five equal splits. For each fold, we considered 4 splits as the training set to which we fitted the model, and the remaining split as the test set on which we made predictions. In the results we show, we use the smallest number of latent groups $K$, $J$, and $L$ for which the prediction accuracy had already reached saturation values. These values were $K=5$, $L=5$ for the tensorial model, and $J=2$, $L=2$ for the bipartite model.
%
%For each fold, we repeated the fitting processes between 50 and 100 times with different random initializations. The results we present correspond to the average over the results for the five folds.

\begin{table*}
\caption{{\bf Dataset characteristics.} The email networks we consider are complete networks where no-links are treated as links of type $0$, so all potential links in the network are observed links. The drug-drug interaction network is a sparse dataset, where we only have information about $1.4\%$ of the links. Each observed drug-drug interaction can be of three types: antagonistic - ANT, additive - ADD, or synergistic - SYN.}
\begin{tabular}{@{\vrule height 10.5pt depth0pt width0pt}|l|c|c|c|c|c|c|}
\hline
Dataset & Types of links $R$ & \#Nodes & \#Layers & \#Observables & Fraction observed & \#Observed \\ 
\hline
Email Unit 1 & $\{0,1\}$ & 104 & 365 & 1,954,940 & 100\% & $|G^O|_1$ = 20,807 \\
\hline
Email Unit 2 & $\{0,1\}$ & 114 & 365 & 2,350,965 & 100\% & $|G^O|_1$ = 27,180 \\
\hline
Email Unit 3 & $\{0,1\}$ & 116 & 365 & 2,434,550 & 100\% & $|G^O|_1$ = 23,979 \\
\hline
Email Unit 4 & $\{0,1\}$ & 118 & 365 & 2,519,595 & 100\% & $|G^O|_1$ = 17,508 \\
\hline
Email Unit 5 & $\{0,1\}$ & 141 & 365 & 3,602,550 & 100\% & $|G^O|_1$ = 23,923 \\
\hline
Email Unit 6 & $\{0,1\}$ & 161 & 365 & 4,701,200 & 100\% & $|G^O|_1$ = 20,790 \\
\hline
Email Unit 7 & $\{0,1\}$ & 225 & 365 & 9,198,000 & 100\% & $|G^O|_1$ = 60,238 \\
\hline
Drug-drug interactions & \{ANT, ADD, SYN\} & 69 & 85 & 199,410 & 1.37\% & $|G^O|_{ANT}$ = 385 \\ \cline{1-6}
Drug-drug interactions & \{ANT, non-ANT\} & 69 & 85 & 199,410 & 1.37\% & $|G^O|_{ADD}$ = 1,543 \\ \cline{1-6}
Drug-drug interactions & \{SYN, non-SYN\} & 69 & 85 & 199,410 & 1.37\% & $|G^O|_{SYN}$ = 863 \\
\hline
\end{tabular}
\label{tab:data}
\end{table*}

\subsection{Baseline models}

%Because to our knowledge there is no previous work that tries to solve the general problem we formulate in this manuscript, in order to assess the performance of the tensorial and bipartite models, 

We compare our models to three different baselines. The naive baseline takes into account all the observations of a link $(i,j)$ in the training set. Then, it makes predictions for the unobserved link types $r_{ijl}$ based on the fraction of times link $(i,j)$ has been observed to be of type $r$ in the training set 
\begin{equation}
P_{\rm naive}[r_{ijl}=r]=\frac{\sum_{s|(ijs) \in G^O} \delta{r,r_{ijs}}}{N_{ij}}~,
\label{naive}
\end{equation}
where $N_{ij}$ is the number of times the $(i,j)$ is observed in the training set.

%For example, if the training set contains 100 observations of a binary interaction between nodes $i$ and $j$ (in 100 different layers), and 40 of them are of type 1, $P_{\rm naive}[r_{ijl}=r]=0.4$ for all layers $l$ in which the type of link $(i,j)$ is not observed.

The other two baselines help us to assess to what extent our models are able to exploit correlations between different layers. First, the independent-layer naive estimates the probability of link $(i,j,l)$ being of type $r$ as the fraction of links of type $r$ observed in layer $l$,
\begin{equation}
P_{\rm naive - IL}[r_{ijl}=r]=\frac{\sum_{(kn)|(knl) \in G^O} \delta{r,r_{knl}}}{N_l}~,
\label{naive-1L}
\end{equation}
where $N_l$ is the number of links of any type observed in layer $l$ of the training set.
Second, we consider an independent-layer mixed-membership stochastic block model for each layer so that,
\begin{equation}
P_{\rm IL}[r_{ijl}=r]= \sum_{\alpha\beta}\theta_{i\alpha}^l\theta_{j\beta}^l p_{\alpha\beta}^l(r)~,
\label{MMSBM-1L}
\end{equation}
where the superindex $l$ denotes that each layer has its own set of parameters. As in the tensorial and bipartite layered models, parameters are subject to the constraints $\sum_{\alpha}\theta^l_{i\alpha}=1,\, \forall l$ and $\sum_r p^l_{\alpha\beta}(r)=1\, \forall l$.  The parameters for this model are obtained using the same method as in the tensorial and bipartite mixed-membership models, but considering each layer separately (see also \cite{godoy-lorite16b}).

\subsection{Email networks}

We first consider the ability of each model and baseline to predict unobserved links in the email networks listed in Table \ref{tab:data}. To assess the performance of each model for each network, we calculate the area under the ROC curve (AUC), the precision, and the recall in 5-fold cross-validation experiments (see Appendix~\ref{app:details} for details). The AUC measures how well a model separates active links (type-1, for which there is communication between the individuals) from inactive links (type-0, with no communication). In particular, it measures the frequency with which an active unobserved link is assigned a higher probability to be active than an inactive unobserved link. Precision accounts for the fraction of links predicted to be active that are indeed active. Recall gives the fraction of active links that are predicted to be active. To calculate both precision and recall, we need to set a threshold $T$ that allows to map probabilities $P[r_{ijl}=1]$ into a binary variable. We do it as follows: if $P[r_{ijl}=1]\geq T$ then the model predicts that $r_{ijl}=1$, otherwise it predicts $r_{ijl}=0$.
In what follows, we choose $T$ as the density of active links in the training set. The reasoning behind this decision is that, because we are splitting data at random into a training and a test set, the test set should have a fraction of active links close to that of the training set.
\footnote{This particular choice of threshold leads, when models are properly calibrated in a frequentist sense, to both precision and recall having very similar values (see Supplementary Information).}

In Fig.~\ref{fig:emails-Performance}, we show that the bipartite link-based model outperforms the tensorial and baseline models in all metrics (see Fig.~S1 for all other email units).
\begin{figure}[]
	\centerline{
\includegraphics*[width=.5\textwidth]{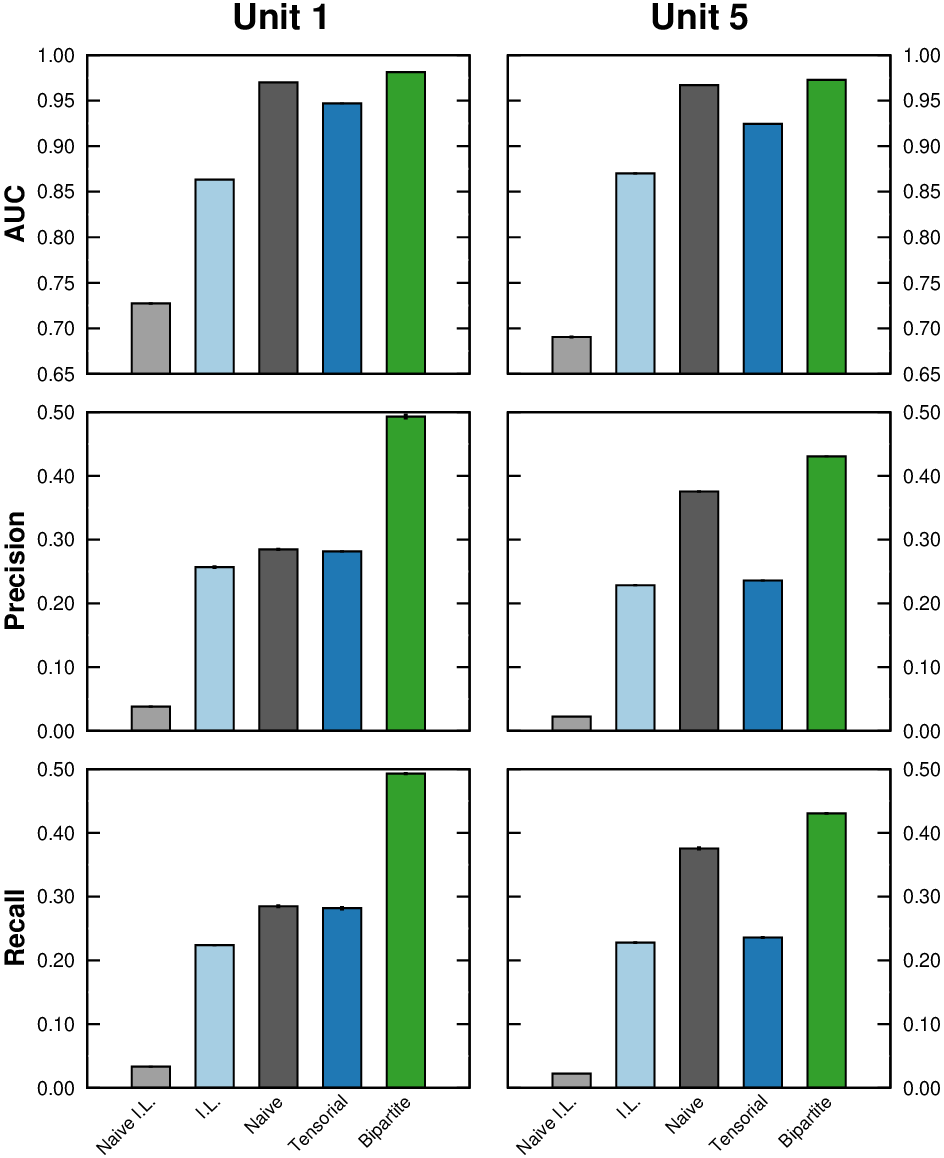}
}
	\caption{{\bf Predictive performance of the models for email networks.}
	{\bf Top:} AUC; {\bf Middle:} Precision; {\bf Bottom:} Recall. Each bar represents the average of the 5-fold cross-validation for a given model (see Appendix~\ref{app:details}). The error bars (shown as a vertical line, which is small and not visible in some cases) represent the standard error of the mean. 
	}
	\label{fig:emails-Performance}
\end{figure}
In these email networks, the AUC is quite high even for the naive baseline because most pairs of individuals never exchange an e-mail and therefore it is easy to predict links for which $r_{ijl}=0$ in all observed layers. The situation changes when we look at precision and recall, which clearly show that the bipartite model is consistently and significantly superior at predicting links that are active. Somewhat surprisingly, we also find that the tensorial node-based model gives slightly lower values than the naive baseline model. 
The explanation lies in the fact that, contrary to both the naive baseline and the bipartite models, the tensorial model focuses primarily on nodes rather than on links and is thus less likely to account for the fact that many pairs of nodes in the network never communicate. More precisely, the probabilities assigned by the tensorial model depend on the product of the membership of the involved nodes, and these memberships are rarely equal to zero. Hence, according to the tensorial model most links have a non-zero probability of existing, including those that are inactive for all observations in the training set.
%This undesired effect leads to many type-0 links obtaining a higher probability than they should, and thus to a lower AUC score.

To further investigate the workings of each approach, we analyze whether they are properly calibrated in a frequentist sense, that is whether the fitted models are able to reproduce statistical features of the training dataset \cite{gneiting07}. In particular, we consider the marginal and probabilistic calibration of all models. A model is probabilistically calibrated if events to which the model assigns a probability $p$ are observed with frequency $p$ \cite{gneiting07}. In our case, a model is calibrated if a fraction $p$ of the links for which $P[r_{ijl}=1]=p$ actually exist. A model is marginally calibrated if, on average, each type of event is assigned a probability that is equal to the actual frequency of such events in the training set. In our case, a model is calibrated if the mean $P[r_{ijl}=1]$ assigned to links coincides with the density of the observed network \cite{gneiting07}.

\begin{figure}[]
	\centerline{
\includegraphics*[width=.5\textwidth]{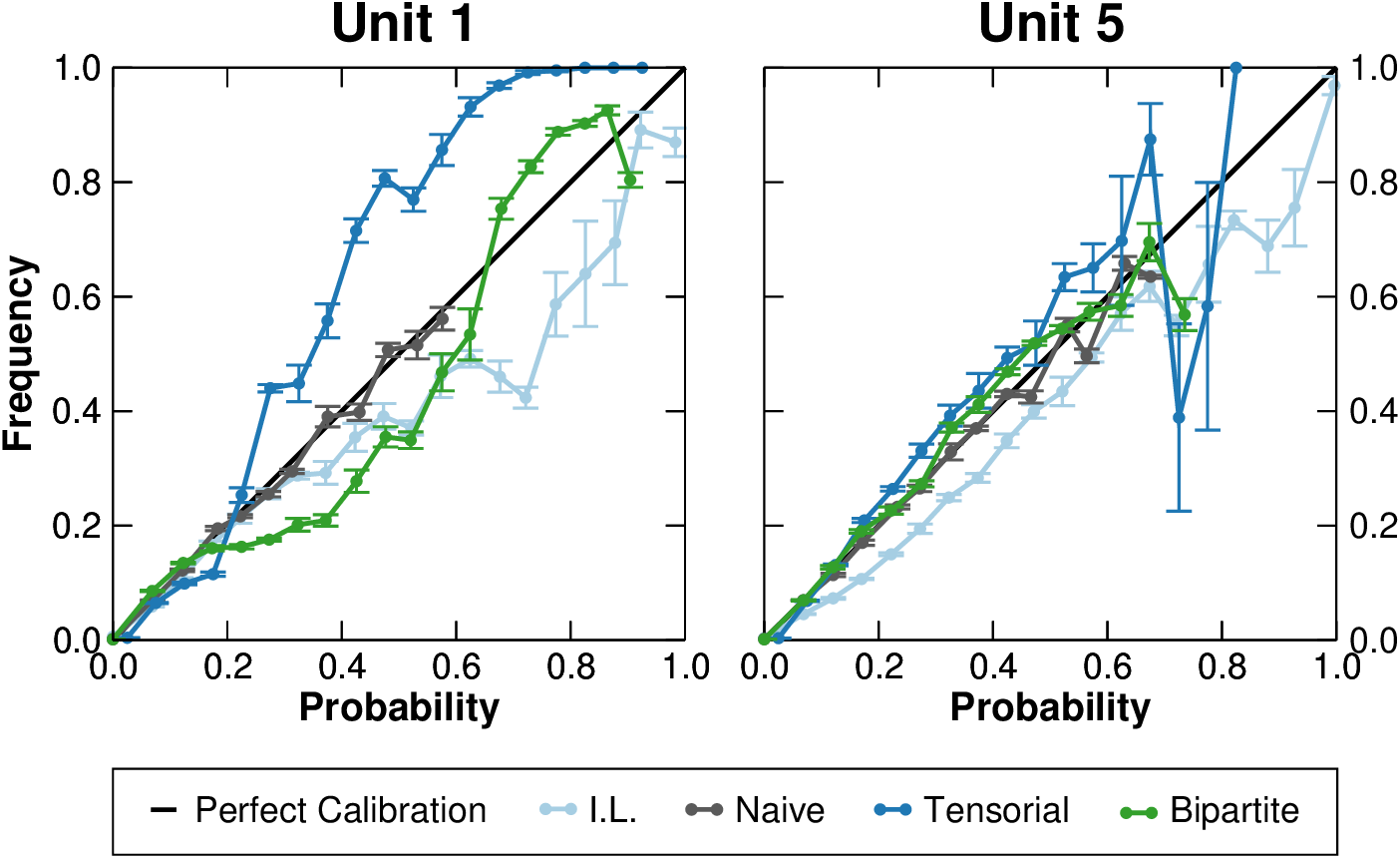}
}
	\caption{{\bf Probabilistic calibration of the models for email networks.}
	Each point in each line represents the average of the 5-fold cross-validation for a given model, with error bars representing the standard error of the mean. The Naive I.L. model is not included as it only assigned tiny probabilities that resulted in a single data point near the origin.
	}
	\label{fig:emails-Calibration}
\end{figure}

\begin{figure}[]
	\centerline{
\includegraphics*[width=.3\textwidth]{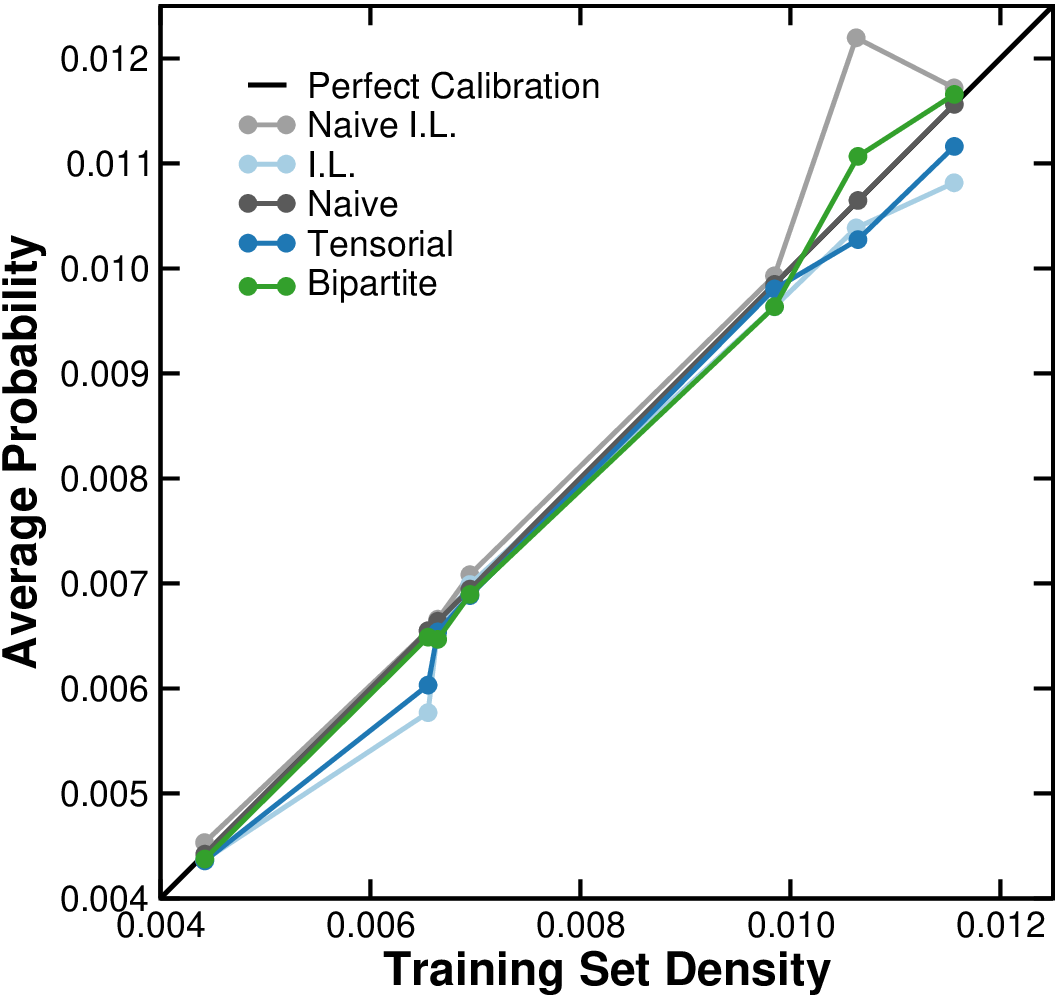}
}
	\caption{{\bf Marginal calibration of the models for email networks.}
	Each line corresponds to a different model, and each point in a line corresponds to a different email network (see Table \ref{tab:data}). Each point represents the average over the 5-fold cross-validation for a given model. Error bars are smaller than symbols.
	}
	\label{fig:emails-MarginalCalibration}
\end{figure}

In Fig.~\ref{fig:emails-Calibration}, we show that all models are relatively well calibrated probabilistically (higher probabilities correspond to higher frequencies), although the calibration is noticeably worse for the network obtained for Unit 1 (see Fig.~S2 for the remaining units). In general, the bipartite model is better calibrated than the tensorial model, which is consistent with the higher predictive accuracy of the bipartite, link-based model. Perhaps surprisingly, the naive baseline model appears to have an even better probabilistic calibration across all units. Figure~\ref{fig:emails-MarginalCalibration} also shows that all models are marginally calibrated.

In light of these observations, the difference in performance between bipartite and naive models must come from the fact that the bipartite model is able to detect temporal patterns that are relevant for the prediction of active links. Indeed, we find that for all the email networks we consider, temporal layers (days) are classified either as week days or as weekend days (and holidays), so that it is more likely for any link to be active on a week day. Interestingly, this is all the temporal information required to be able to accurately predict whether a specific link is going to be active or not on a certain day \footnote{Note that our results do not depend on the number of latent dimensions allowed for the temporal layers L, since for $L>2$ we also find that temporal layers have $\eta_{\ell \gamma}\neq 0$ only for two latent groups $\gamma$}.

%Note, however, that the highest probabilities assigned by this model are around 0.6, which indicates that the model is never really certain about the existence of a link. This is probably due to the inability of the baseline model to distinguish between days (working versus non-working), and explains why the bipartite model (which does group days and assigns probabilities close to 1 in some cases) has better predictive performance 

%

\subsection{Drug-drug interactions in cancer}

\begin{figure}[t!]
	\centerline{
\includegraphics*[width=.5\textwidth]{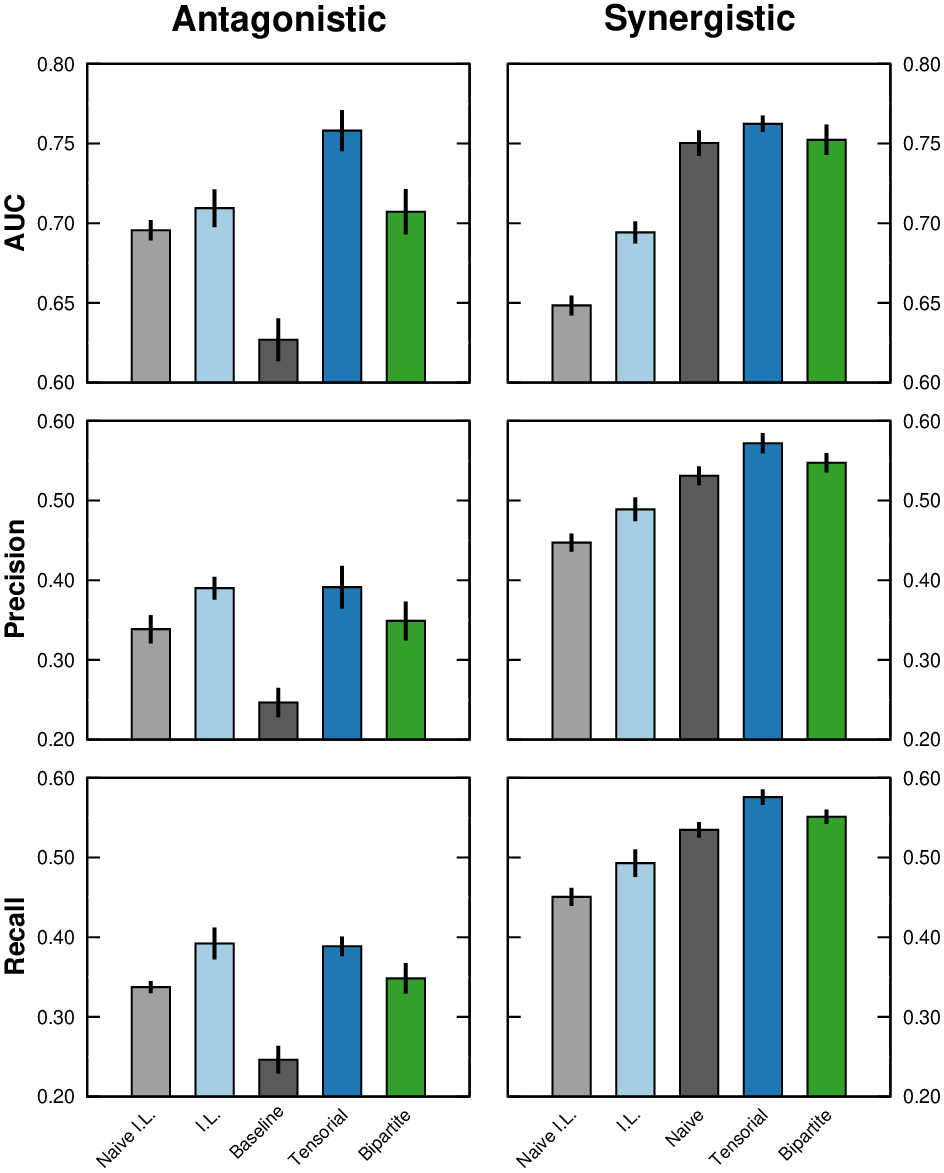}
}
	\caption{{\bf Predictive performance of the models for drug-drug interaction networks.}
	{\bf (a)} AUC statistic. {\bf (b)} Precision. {\bf (c)} Recall. Each bar represents the average of the 5-fold cross-validation for a given model, with error bars (shown as a vertical line for clarity) representing the standard error of the mean.
	}
	\label{fig:drugs-3C-Performance}
\end{figure}

\begin{figure}[t!]
	\centerline{
\includegraphics*[width=.5\textwidth]{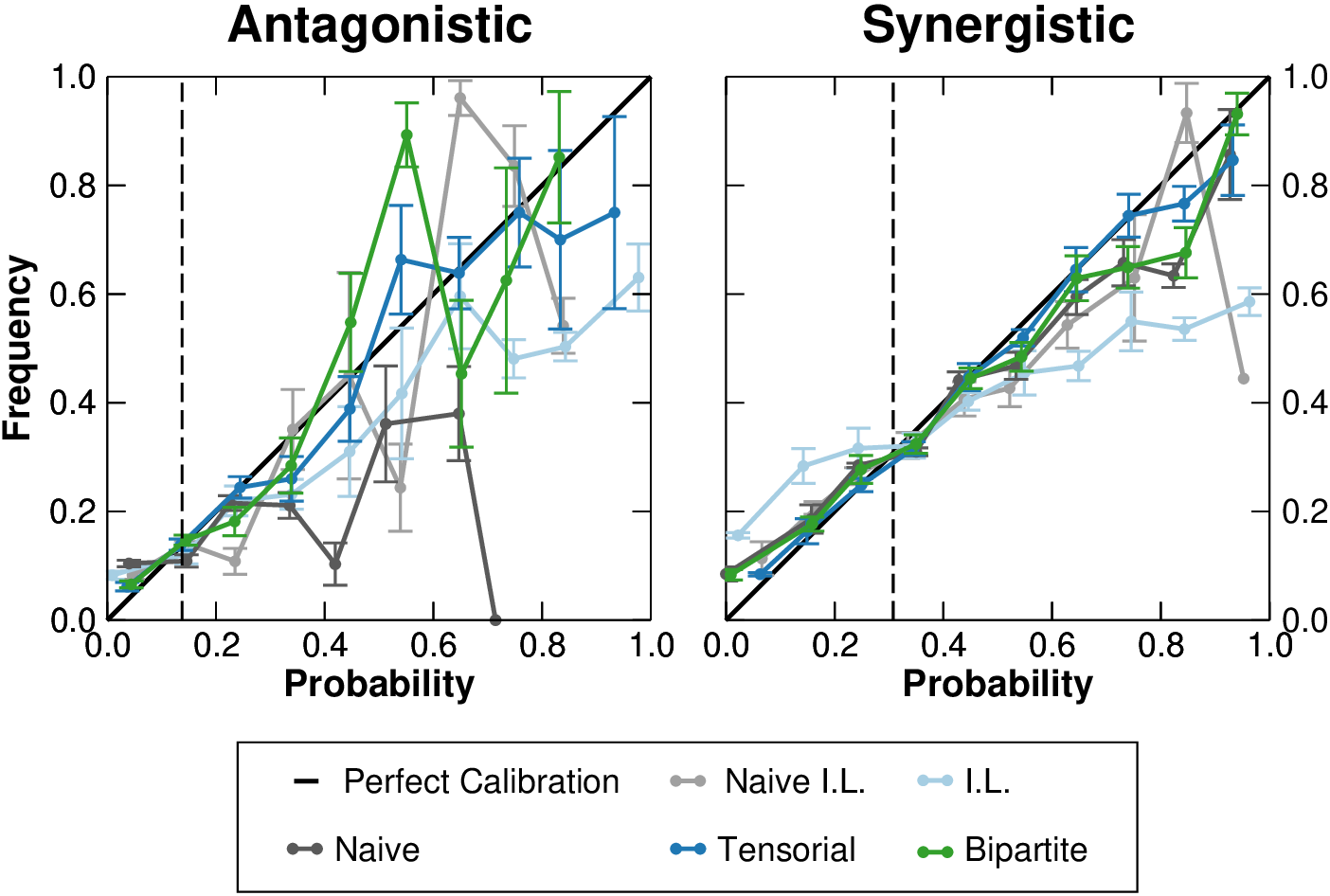}
}
	\caption{{\bf Probabilistic calibration of the models for drug-drug interaction networks.}
	Each point in each line represents the average of the 5-fold cross-validation for a given model. Error bars represent the standard error of the mean. The vertical dashed lines show the density of each training set.
	}
	\label{fig:drugs-3C-Calibration}
\end{figure}

\begin{figure}[t!]
	\centerline{
\includegraphics*[width=.3\textwidth]{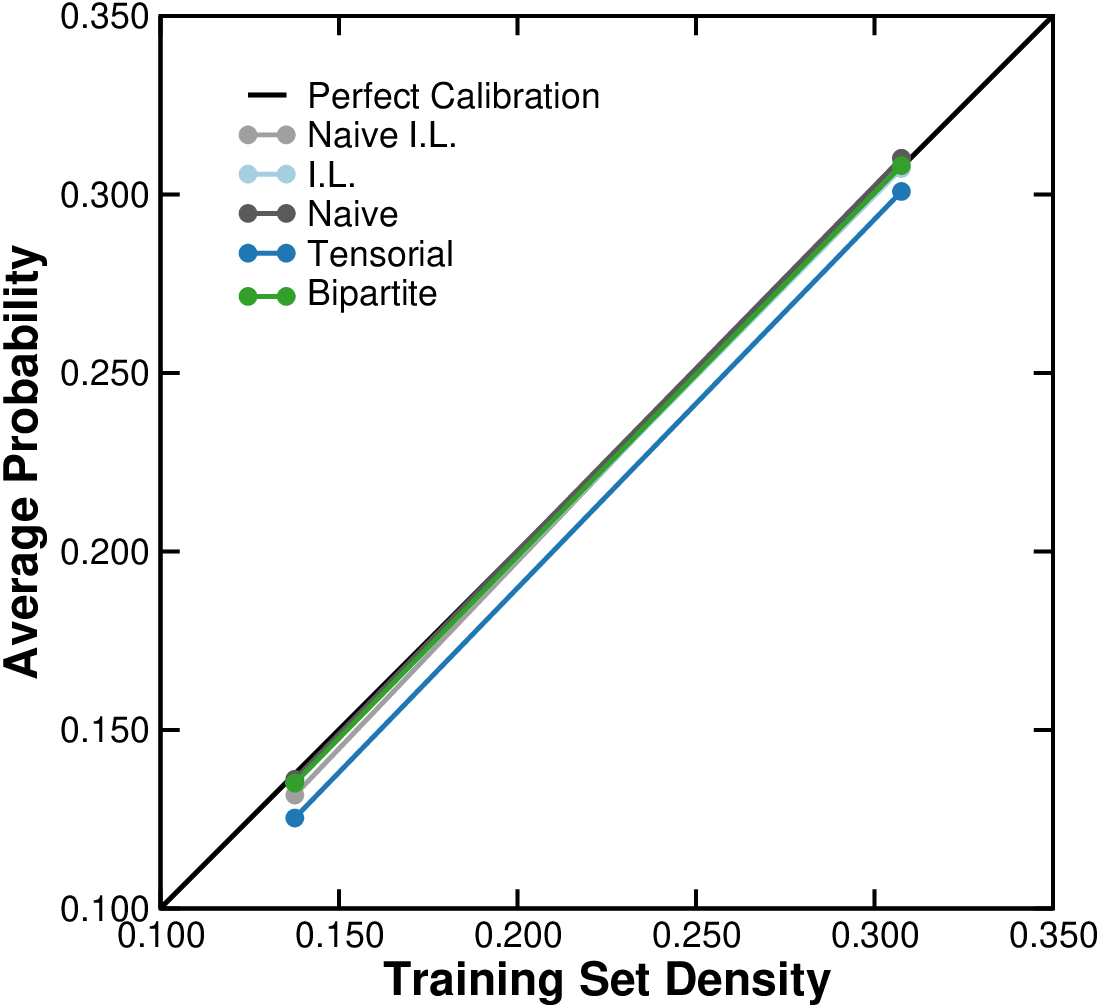}
}
	\caption{{\bf Marginal calibration of the models for drug-drug interaction networks.}
	For each of the models we consider (see legend) we plot the average probability for links being of a certain type (antagonistic or synergistic) with respect to the density of links of that type in the training set. Each point represents the average over the 5 training sets for a given model. Error bars are smaller than symbols.
	}
	\label{fig:drugs-3C-MarginalCalibration}
\end{figure}

Links in the drug-drug interaction network are of three different types: synergistic, antagonistic, and additive; we trained the models considering the three types of interactions. However, because the interesting question is whether synergistic or antagonistic interactions can be predicted, we evaluated the performace of each model for each task of these two tasks. For instance, to evaluate the accuracy of a model at predicting synergistic interactions, we binarized model predictions into synergistic and non-synergistic. We then computed the metrics over this binary outcome as we did for e-mail networks  (Figs. \ref{fig:drugs-3C-Performance}, \ref{fig:drugs-3C-Calibration}, and \ref{fig:drugs-3C-MarginalCalibration}; Fig.~S3 shows that all of the results below are qualitatively similar when training our models on networks with only two types of interactions: synergistic/non-synergistic or antagonistic/non-antagonistic) \footnote{Due to the sparsity of observations in these networks many interactions were never observed in the training sets, and thus no group memberships could be assigned to the links ($e_{ij}$) corresponding to those interactions. We solved this cold start problem by, at each iteration, assigning them the average membership of the observed interactions $\zeta_{e}={\left\langle \zeta_f \right\rangle}_{f \in G^O}$. Analogously, if a node $i$ had no observed interactions in the training set, at each iteration we set its membership vector as the average of membership vectors for nodes with observed interactions $\theta_i={\left\langle \theta_k \right\rangle}_{k \in G^O}$.}.

Contrary to what we observed for the email networks, we find that the tensorial model performs better than the bipartite model. Our results thus suggest that for this dataset, grouping nodes (drugs) into groups summarizes more parsimoniously the information relevant for prediction. This is consistent with previous findings that show that mechanisms of action and target pathways of drugs are related to the effect they display when combined with other drugs, an information that is best captured by node memberships than by link memberships \cite{guimera13}.

%We observe, that AUC values are lower than those obtained for e-mail networks because the spasity of the dataset makes it a harder problem.

Interestingly, we observe differences in performance at detecting antagonistic and synergistic interactions. For the synergistic interaction network, we find that the tensorial model consistently outperforms the bipartite and baseline models in all metrics (AUC, precision and recall), although its marginal calibration is slightly worse that that of the other models.
%, since the tensorial model assigns  probabilities to synergistic interactions that are on average lower than the density of the training set.
%
For antagonistic interactions, the tensorial model also performs better than the bipartite and baseline models in terms of AUC. However, the tensorial model has a precision and recall that are similar to those of the independent-layers baseline model. The generalized decrease in precision and recall with respect to synergistic network does not come as a surprise since none of the models is perfectly calibrated for probabilities lower than the density of the training set (Fig.~\ref{fig:drugs-3C-Calibration}). In fact, we observe that the fraction of antagonistic interactions for which $P[r_{ijk}=1]<T$ is larger than desired. As a result, some antagonistic interactions are counted as non-antagonistic interactions in terms of precision and recall. This effect is exhacerbated by the fact that, due to the sparsity of the network, a large fraction of interactions are assigned low probability values by all of the models.

The fact that the independent-layer model has prediction and recall values similar to those of the tensorial model can be explained by the fact that antagonistic interactions are more localized to specific layers than synergistic interactions are (see Fig.~S6). 
This situation makes it easier for the independent-layer baseline model to make more accurate predictions for these layers. Note however, that if more information on antagonistic interactions was available, the performance of the tensorial model would likely be comparable to that of the synergistic case.

%With observed antagonistic interactions being much more sparse than synergistic ones, it is harder for any model to assign an interaction a large probability of being or not being antagonistic. While this does not necessarily affect the AUC because all the assigned probabilities are lower and their relative rankings can still be accurate, it must definitely have an effect on precision and recall as these metrics highly depend on the probability threshold $T$. If the models are assigning lower probabilities, a higher fraction of the predictions will fall below the threshold and will increase the amount of false negatives, thus 

%Figure~\ref{fig:drugs-2C-Performance}, we show the predictive performance of the models for both networks. As we did for the email networks, we compare them to the baseline model, and to the independent-layers version of the baseline and the tensorial models. Overall, predictions in the synergistic network are consistently and significantly better than in the antagonistic network, which we surmise is due to the fact that the dataset contains more synergistic than antagonistic interactions.

%
%
%In terms of AUC, the tensorial model is the most predictive for both networks. In terms of precision and recall, the tensorial model is also clearly superior to all others for the synergistic network. For the antagonistic network, both the tensorial and the independent-layers tensorial models give almost identical results. Overall, then, and unlike in email networks, the tensorial model is consistently superior to the bipartite model.

%With regards to probabilistic calibration, all models except the independent-layers tensorial model are well calibrated for the synergistic network (Fig.~\ref{fig:drugs-2C-Calibration}). Calibration is not so good for the antagonistic network, which, again, is a consequence of the fact that the amount of observed antagonistic interactions is smaller than the amount of observed synergistic interactions. In any case, and considering the error bars in the case of the antagonistic network, the models are well generally calibrated to the data also in this case. The models are also marginally calibrated (Fig.~\ref{fig:drugs-2C-MarginalCalibration}) \textbf{[OBS: deixem aquesta figura aixi, o busquem alguna manera millor de representar-la?]}.
%

\section{Discussion}

We have presented two mixed-membership multi-layer network models that can be applied to any multi-layer networks, with layers representing temporal snapshots of the interactions or different contexts for the interactions. By extending the mixed-membership paradigm to the layers themselves, and by not making any prior assumption about them, our models can detect and take advantage of inter-layer correlations in the network of interactions to make better predictions. As a result, both our multi-layer models outperform the baseline models in almost all the studied cases, except for the cases in which information is too sparse for the multi-layer model to recover unobserved interactions with precision.

Importantly, none of the models we present---the tensorial node-based model or the bipartite link-based model---is intrinsically better than the other; however they can hold clues as to the mechanisms that are predictive of interaction types. Our results precisely illustrate this fact. We find that the bipartite model works better for email networks in which the communication between pairs of users (links), rather than the users themselves, together with their temporal evolutions are the relevant description unit for prediction. This could be due to the fact that we are analyzing communication at a rather small scale (people working within the same unit of an organization), and it is possible that a node-based model could be better for communication between users at a larger scale. Moreover, as the network grows the number of $\theta$ parameters for the tensorial model scales linearly with the number of nodes, whereas the number of $\eta$ parameters for the bipartite model scales quadratically. In really big networks it is then plausible that the tensorial becomes more parsimonious.
  
Conversely, our results show that for the drug-drug interaction network the relevant unit of description are drugs (nodes). This is consistent with the fact that the mechanism of action/target that determines how a drug will interact with another one; this information is encapsulated in the node (and its observed interactions). The use of the interactions of nodes  in different cancer types (layers) boosts our ability to predict the type of type-dependent interactions more precisely. In contrast, the description of these networks in terms of drug-pair interactions completely misses the drug-specific information that is relevant for prediction in this context. 

Our results unambiguously show that using the information of the interactions on other layers helps obtain better models. Remarkably, the flexibility of the models we propose make this approach suitable to analyze multi-layer networks in any context. A natural step to further improve the model and prediction accuracy would be to include auxiliary data (i.e. metadata such as node or link attributes) into the modeling process. This problem has just started being explored in the literature \cite{newman16,hric16,sapienza17}, so there is no general framework on how to introduce auxiliary data into the inference process yet. Nonetheless, recent results show that single-layer mixed-membership models are suitable models to incorporate specific types of auxiliary data into the inference process without adding methodological complexity \cite{cobo17}, thus opening the window to developing general inference frameworks that consider different types of metadata also in multilayer contexts.

\begin{acknowledgments}
This work was supported by the Spanish Ministerio de Economia y Comptetitividad (MINECO) Grant FIS2016-78904-C3-1-P, and European Union FET Grant 317532 (MULTIPLEX). The authors acknowledge the computer resources at MinoTauro and the technical support provided by the Barcelona Supercomputing Center (FI-2015-3-0024, FI-2016-3-0026). The authors acknowledge AstraZeneca UK Limited, Sanger and Sage Bionetworks-DREAM for providing the cancer drug interaction data.
\end{acknowledgments}

% APPENDICES

\setcounter{equation}{0}
\renewcommand\theequation{A.\arabic{equation}}
\appendix
\section{Derivation of the expectation maximization equations for the T-MBM}
\label{app:tensorial}
In the tensorial mixed-membership stochastic block model, we assign membership vectors $\theta_{i\alpha}$, $\eta_{\ell \gamma}$ to each node $i$ and each layer $\ell$, respectively. These membership vectors are properly normalized, therefore represent the probability that each node/layer belongs to a specific node/layer group:
\begin{equation}
\label{eqapp:norm-pm}
\forall i: \sum_{\alpha=1}^K \theta_{i\alpha} = 1 \, ,
\quad \forall \ell: \sum_{\gamma=1}^L \eta_{\ell \gamma} = 1 \, .
\end{equation}

Because we consider that links cant take different values $r \in R$, to ensure that each observed interaction has probability 1 of receiving any rating, we normalize probability matrices $p_{\alpha \beta \gamma}(r)$ 
\begin{equation}
\label{eqapp:norm-pr}
\forall \alpha, \beta, \gamma: \sum_{r \in R} p_{\alpha \beta \gamma}(r) = 1 \, .
\end{equation}
Note that if $R=\{0,1\}$, then $ p_{\alpha \beta \gamma}(0)=1- p_{\alpha \beta \gamma}(1)$.

We maximize the likelihood~\eqref{eq:likeli-t} as a function of $\thetas, \etas, \ps$ using an expectation maximization (EM) algorithm. We start with a standard variational and use Jensen's inequality $\log \bar{x} \ge \overline{\log x}$ in order to transform the logarithm of a sum into a sum of logarithms 
\begin{align}
\log & P(G^O | \thetas, \etas, \ps ) 
= \sum_{(ij\ell) \in G^O} \log \sum_{\alpha\beta\gamma} \theta_{i\alpha} \theta_{j\beta} \eta_{\ell \gamma} p_{\alpha \beta \gamma}(r_{ij\ell}) \nonumber \\
&= \sum_{(ij\ell) \in G^O} \log \sum_{\alpha\beta\gamma} \omega_{ij\ell}(\alpha\beta\gamma) \,\frac{\theta_{i\alpha} \theta_{j\beta} \eta_{\ell \gamma} p_{\alpha \beta \gamma}(r_{ij\ell})}{\omega_{ij\ell}(\alpha\beta\gamma)} 
\nonumber \\
&\ge \sum_{(ij\ell) \in G^O} \sum_{\alpha\beta\gamma} \omega_{ij\ell}(\alpha\beta\gamma) \log \frac{\theta_{i\alpha} \theta_{j\beta} \eta_{\ell \gamma} p_{\alpha \beta \gamma}(r_{ij\ell})}{\omega_{ij\ell}(\alpha\beta\gamma)} \, . 
\label{eqapp:final-t}
\end{align}
Here we have introduced the auxiliary variable $\omega_{ij\ell}(\alpha\beta\gamma)$, which is the estimated probability that a given link's type $r_{ij\ell}$ is due to $i$, $j$ and $\ell$ belonging to groups $\alpha$, $\beta$ and $\gamma$ respectively. Note in the expression above, equality holds when 
\begin{equation}
\label{eqapp:update-omega-t}
\omega_{ij\ell}(\alpha\beta\gamma) 
= \frac{\theta_{i\alpha} \theta_{j\beta} \eta_{\ell \gamma} p_{\alpha \beta \gamma}(r_{ij\ell})}{\sum_{\alpha'\beta'\gamma'} \theta_{i\alpha'} \theta_{j\beta'} \eta_{\ell \gamma'} p_{\alpha' \beta' \gamma'}(r_{ij\ell})} \, . 
\end{equation}
This is precisely the equation for the expectation step.

For the maximization step, we derive update equations for the parameters $\thetas, \etas, \ps$ by taking derivatives of the log-likelihood~\eqref{eqapp:final-t}. Including Lagrange multipliers for the normalization constraints~\eqref{eqapp:norm-pm}, we obtain for $\theta_{i\alpha}$
\begin{equation}
\label{eqapp:update-theta-t}
\theta_{i\alpha} = \frac{\sum_{j\ell \in \partial i} \sum_{\beta\gamma} \omega_{ij\ell}(\alpha\beta\gamma)}
{\sum_{j\ell \in \partial i} \sum_{\alpha\beta\gamma} \omega_{ij\ell}(\alpha\beta\gamma)} 
= \frac{\sum_{j\ell \in \partial i} \sum_{\beta\gamma} \omega_{ij\ell}(\alpha\beta\gamma)}{d_i} \, ,
%\mid (u,i) \in R 
\end{equation}
where $\partial i = \{ j,\ell | (ij\ell) \in G^O \}$ and $d_i=|\partial i|$ is the degree of node $i$ in all the layers for any type of link. Similarly, for $\eta_{\ell\gamma}$ we obtain
\begin{equation}
\label{eqapp:update-eta-t}
\eta_{\ell\gamma} = \frac
{\sum_{ij \in \partial \ell} \sum_{\alpha\beta} \omega_{ij\ell}(\alpha\beta\gamma)}
{\sum_{ij \in \partial \ell} \sum_{\alpha\beta\gamma} \omega_{ij\ell}(\alpha\beta\gamma)} 
= \frac{\sum_{ij \in \partial \ell} \sum_{\alpha\beta} \omega_{ij\ell}(\alpha\beta\gamma)}{d_{\ell}} \, ,
\end{equation}
where $\partial \ell = \{ i,j | (ij\ell) \in G^O \}$ and $d_{\ell}=|\partial \ell|$ is the number of observed links of any type in layer $\ell$.

Finally, including a Lagrange multiplier for~\eqref{eqapp:norm-pr}, we have for $p_{\alpha\beta\gamma}(r)$
\begin{equation}
\label{eqapp:update-pr}
p_{\alpha\beta\gamma}(r)= \frac
{\sum_{(ij\ell) \in G^O | t_{ij\ell}=t} \omega_{ij\ell}(\alpha\beta\gamma)}
{\sum_{(ij\ell) \in G^O} \omega_{ij\ell}(\alpha\beta\gamma)} \, .
\end{equation}

\section{Derivation of the expectation maximization equations for the B-MBM}
\label{app:bipartite}
%Therefore, we infer the values of the parameters $\hatzetas, \hatetas, \hatps$ that maximize this likelihood using the expectation-maximization method. 
As in the tensorial model, we assign normalized membership vectors $\zeta_{e\alpha}$, $\eta_{\ell\gamma}$ to links and layers, respectively. We also consider probability matrices $p_{\alpha\gamma}(r)$ that are as well normalized $ \sum_{r\in R} p_{\alpha\gamma}(r) = 1 $).

In order to maximize the likelihood, we again use Jensen's inequality to transform the the logarithm of a sum into a sum of logarithms and introduce an auxiliary variable $\phi_{e\ell}(\alpha, \gamma)$: 
\begin{align}
\log & P(G^O | \zetas, \etas, \ps ) 
= \sum_{(ij\ell) \in G^O} \log \sum_{\alpha\gamma} \zeta_{e\alpha} \eta_{\ell \gamma} p_{\alpha \gamma}(r_{e\ell}) \nonumber \\
&= \sum_{(ij\ell) \in G^O} \log \sum_{\alpha\gamma} \phi_{e\ell}(\alpha, \gamma) \,\frac{\zeta_{e\alpha} \eta_{\ell \gamma} p_{\alpha \gamma}(r_{e\ell})}{\phi_{e\ell}(\alpha, \gamma)} 
\nonumber \\
&\ge \sum_{(e\ell) \in G^O} \sum_{\alpha\gamma} \phi_{e\ell}(\alpha, \gamma) \log \frac{\zeta_{e\alpha} \eta_{\ell \gamma} p_{\alpha \gamma}(r_{e\ell})}{\phi_{e\ell}(\alpha, \gamma)} \, . 
\label{eqapp:links-final-b}
\end{align}
where again the equality holds when
\begin{equation}
\label{eqapp:links-update-phi}
\phi_{e\ell}(\alpha, \gamma) 
= \frac{\zeta_{e\alpha} \eta_{\ell \gamma} p_{\alpha \gamma}(r_{e\ell})}{\sum_{\alpha'\gamma'} \zeta_{e\alpha'} \eta_{\ell \gamma'} p_{\alpha' \gamma'}(r_{e\ell})} \, , 
\end{equation}
giving us the update equation~\eqref{eqapp:links-update-phi} for the expectation step.

For the maximization step, we derive update equations for the parameters $\zetas, \etas, \ps$ by taken derivatives of the log-likelihood~\eqref{eqapp:links-final-b}. Including Lagrange multipliers for the normalization constraints, we obtain 
\begin{equation}
%\label{eq:links-update-theta}
\zeta_{e\alpha} = \frac{\sum_{\ell \in \partial e} \sum_{\gamma} \phi_{e\ell}(\alpha, \gamma)}
{\sum_{\ell \in \partial e} \sum_{\alpha\gamma} \phi_{e\ell}(\alpha, \gamma)} 
= \frac{\sum_{\ell \in \partial e} \sum_{\beta\gamma} \phi_{e\ell}(\alpha, \gamma)}{d_{e}} \, ,
%\mid (u,i) \in R 
\end{equation}
where $\partial e = \{\ell | (e\ell) \in G^O \}$ are the set of layers in which we observe link $e_{ij}$ and $d_{e}=|\partial e|$ is the total number of layers in which we observe link $e_{ij}$. Similarly, 
\begin{equation}
\label{eqapp:links-update-eta}
\eta_{\ell\gamma} = \frac
{\sum_{e \in \partial \ell} \sum_{\alpha} \phi_{e\ell}(\alpha, \gamma)}
{\sum_{e \in \partial \ell} \sum_{\alpha\gamma} \phi_{e\ell}(\alpha, \gamma)} 
= \frac{\sum_{e \in \partial \ell} \sum_{\alpha} \phi_{e\ell}(\alpha, \gamma)}{d_{\ell}} \, ,
\end{equation}
where $\partial \ell = \{ e | (e\ell) \in G^O \}$ and $d_{\ell}=|\partial \ell|$. Finally, including a 
Lagrange multiplier for the normalization of $p_{\alpha\gamma}(r)$, we have
\begin{equation}
\label{eqapp:links-update-pr}
p_{\alpha\gamma}(r)= \frac
{\sum_{(e\ell) \in G^O | r_{e\ell}=r} \phi_{e\ell}(\alpha, \gamma)}
{\sum_{(e\ell) \in G^O} \phi_{e\ell}(\alpha, \gamma)} \, .
\end{equation}

Equations ~\eqref{eqapp:links-update-phi}-\eqref{eqapp:links-update-pr} are solved iteratively with an EM algorithm following the same procedure as in the tensorial model. The bipartite model also scales linearly with the size of the dataset, but in this case the number of parameters of the model is $IK+ML+|G^O| K\cdot L$, where the number of links $I\leq N\cdot(N-1)/2$, thus, even though it increases the number of parameters (number of nodes $N$ is typically smaller than number of links $I$), there is one dimension less to run over all observed links in all layers $|G^O|$.

\section{Experimental details}
\label{app:details}

With regards to the drug-drug interactions dataset, we divided the continuous values of efficiency into three categories (synergistic, additive and antagonistic) by setting two thresholds as suggested in the original experimental data. These thresholds are -20.0 and 20.0, so that interactions with an efficiency lower than -20.0 are classified as antagonistic, those with an efficiency higher than 20.0 are classified as synergistic, and those in between are considered additive \cite{DREAM15}.

For both datasets, we fitted and validated our models using a 5-fold cross-validation scheme. We first divided the data into five equal splits. Then for each fold we considered 4 splits as the training set to which we fitted the model, and the remaining split was kept as the test set on which we made predictions. In order to select the number of latent groups $K$, $J$, and $L$, we used the smallest values for which the prediction accuracy had already reached saturation values. These values were $K=5$, $L=5$ for the tensorial model, and $J=2$, $L=2$ for the bipartite model.

For each fold, we repeated the fitting processes between 50 and 100 times with different random initializations. The results we present correspond to the average over the results for the five folds.

\end{document}